\documentclass[a4paper,showkeys,floatfix,aps,pre,reprint,groupedaddress]{revtex4-1}
\usepackage{amsmath,amssymb}
\usepackage{graphics,graphicx}
\usepackage{dcolumn,bm}
\usepackage{psfrag}
\usepackage{color}
\newcommand{\srm}
{\affiliation{Department of Physics, SRM University - AP,
 Andhra Pradesh - 522502, India}}

\begin{document}

\title{Prediction of imminent failure using supervised learning in fiber bundle model}

\author{Diksha}

\email{diksha_bhatt@srmap.edu.in}
\srm
\author{Soumyajyoti Biswas}

\email{soumyajyoti.b@srmap.edu.in}
\srm
\begin{abstract}
Prediction of breakdown in disordered solids under external loading in a question of paramount importance. 
Here we use a fiber bundle model for disordered solids and record the time series of the avalanche sizes and
energy bursts. The time series contains statistical regularities that not only signify universality in the
critical behavior of the process of fracture, but also reflect signals of proximity to a catastrophic failure. 
A systematic analysis of these series using supervised machine learning can predict the time to failure. Different features of the
time series become important in different variants of training samples. We explain the reasons for such switch over 
of importance among different features. We show that inequality measures for avalanche time series play a crucial role
in imminent failure predictions, especially for imperfect training sets i.e., when simulation parameters of training samples differ considerably from
those of the testing samples. We also show the variation of predictability of the system as the interaction range and
strengths of disorders are varied in the samples, varying the failure mode from brittle to quasi-brittle (with interaction range) and from
nucleation to percolation (with disorder strength). The effectiveness of the supervised learning is best when the samples just enter 
the quasi-brittle mode of failure showing scale-free avalanche size distributions. 
\end{abstract}


\maketitle

\section{Introduction}
A slowly driven disordered material progresses towards catastrophic failure point through a series of intermittent avalanches \cite{wiley_book, ekhard}. In a wide variety of
situations, for example fracture of quasi-brittle materials, these 
avalanches are of multiple scales, even though the applied drive is at a constant rate. This is a widely studied phenomenon that
comes under the term crackling-noise \cite{sethna} and it is seen in a myriad of systems, starting from rock samples under external stress, 
disordered magnetic materials under external magnetic field  \cite{zap97,zapperi98}, driven vortex lines in type II superconductors \cite{larkin79} to the largest scale
of mechanical failure i.e., earthquakes \cite{wiley_book2}. 

Much have been studied on the statistical nature of the response of the disordered materials under slow drive. 
Particularly, a scale free nature of the size distribution of the avalanches, the rate of the events prior to and after 
a large avalanche have been observed widely. It was also shown, through experimental data, numerical models and 
analytical calculations, that the exponent value of the avalanche size distribution depends on very broadly defined
parameters, such as the dimensionality of the system and the range of interaction within the system. This prompted, 
over the last several decades, the recasting of the problem of fracture and breakdown in driven disordered materials in 
terms of critical phenomena (see e.g., \cite{wiley_book,bkc_book,rmp} for reviews). Like in the critical phenomena, for example, in magnetic systems, the exponent values
of different quantities, measured near the critical (breakdown) point are largely insensitive to the details of the system
under study. This has resulted in efforts to capture these emergent phenomena through much simplified models (like the 
Ising model in the magnetic system). The random fiber bundle model is one such crucial attempts, which was introduced in 
the textile industry \cite{pierce}. Later on it came 
to be known as one of the paradigmatic model of fracture of disordered solids \cite{rmp}.

\begin{figure*}
\includegraphics[width=16cm]{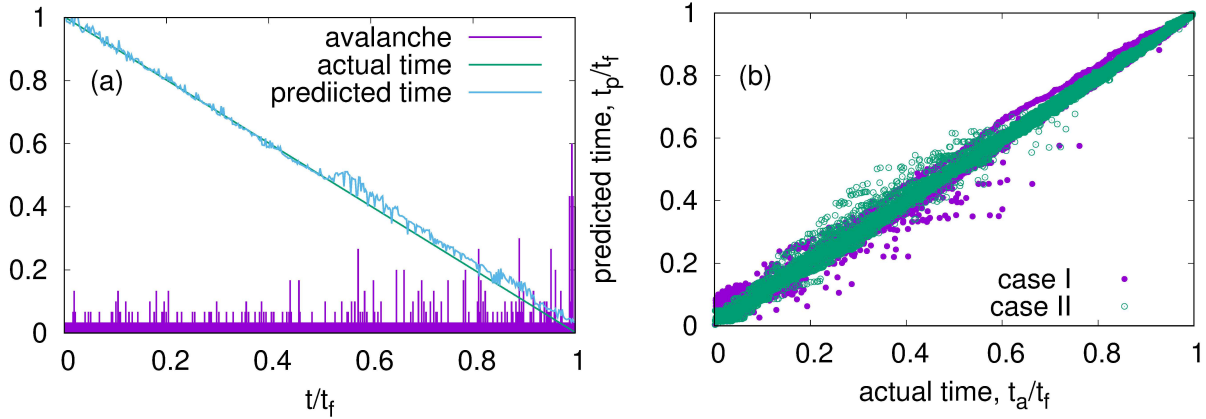}
\caption{Supervised learning based predictions of failure time in the ELS fiber bundle model. (a) The time series of 
(scaled) avalanches are shown, along with the actual remaining time until catastrophic breakdown (straight line) and the
remaining time predicted by the machine learning algorithm (fluctuating line). (b) The actual ($t_a$) and predicted ($t_p$) remaining
times are plotted, scaled by the failure time $t_f$. Case I represented the predictions using a training data where the inequality measures of the avalanche series
were not used and in Case II the inequality measures ($g,h,k$) were used. In both cases, time steps, avalanche size and energy burst sizes were used. A diagonal straight line would be a perfect prediction, however a higher scatter is seen in 
Case I near the origin i.e., closer to failure than in Case II. This is, therefore, a schematic 
demonstration of the importance of the inequality measures as training features. }
\label{first_fig}
\end{figure*}

The fiber bundle model shows a dynamical critical behavior in failure of disordered materials, with the failure modes 
determined by the range of stress release and the strength of the disorder in the model.
It was shown that the fiber bundle model
can reproduce different modes of failure, such as the brittle failure, quasi-brittle failure, ductile failure and so on \cite{failure_mode}.  
 Therefore, in this simple model 
varying just a few parameters different failure modes can be obtained. 
The different failure modes have different characteristics, which are reflected in the time series of the intermittent failures (crackling noise). This, in turn, 
determines the predictability of failure in the models. For example, very low disorder would lead to brittle failure, which is abrupt and hence unpredictable. 
On the other hand, higher disorder leads to increase in predictability \cite{sci_rep1,sci_rep2}. Similarly, range of interaction plays a crucial role. If the damage is fully localized, 
the mode of failure is nucleation driven, hence less predictable.  

It has long been noted that the a non-invasive way of monitoring stability of a disorder sample is to analyze the time series of acoustic emission emitted by it. In particular, the intermittent nature of the acoustic time series and, when available, the data for spatial locations of damages mirror the growing instability in the system i.e., the precursor for the onset of a catastrophic event \cite{ekhard,wiley_book}. There have been empirical rules regarding the statistical regularities of such damages that can estimate the time remaining for the onset of breakdown, which are applied to stability analysis of mechanical structures, to predict snow avalanches \cite{snow}, cliff collapse \cite{cliff}, creep ruptures \cite{exp1,exp2}, power grid outages \cite{grid} and creating earthquake hazard maps \cite{map1,map2}. Similar analysis were also done for numerical models, staring from molecular dynamics simulations to threshold activated cellular automata based models \cite{wiley_book}. The fiber bundle model falls in the second category, and its advantage is that at least in the mean field limit, some of these empirically observed rules are analytically derived as well. 

These rule based predictions for onset of failure, nevertheless, remain of limited success, given the sample to sample fluctuations and limited number of samples in experiments (see e.g., \cite{exp3}). On the other hand, extracting statistical trends from multidimensional data sets is precisely for what machine learning methods are used. Therefore, assuming that the time series of `crackling noise' holds the key for imminent failures, it is natural to use (supervised) machine learning algorithms for predictions of failure using such time series and their various properties as attributes (see e.g., \cite{leduc,sci_rep2} for some recent efforts).

In this work, we use the time series of avalanches (number of fibers breaking following a stress increase) and the corresponding energy bursts, to predict 
imminent failure in the system. For this we use a supervised machine learning algorithm, specifically the random forest algorithm, and utilize the time series
and its different features. A supervised learning method uses a set of samples to train the algorithm, which can them be used for predicting imminent failure
in other nominally similar samples not previously seen by the algorithm. However, in experiments, this nominal similarity (for example in system sizes and 
disorders) might not be perfect. First we show that this difficulty can be overcome by using several inequality measures (such as the Gini \cite{gini} and the 
Kolkata indices \cite{kolkata}), which were recently shown \cite{k_pre} to exhibit remarkable regularity in the failure dynamics, in the training of the algorithm. We then
show how the prediction accuracy of the algorithm differs as the mode of failure is changed by varying the strength of disorder and interaction range in the model.
The mean-field version of the model with intermediate failure range (quasi-brittle mode) turns out to be most predictable.

\begin{figure}
\includegraphics[width=9cm]{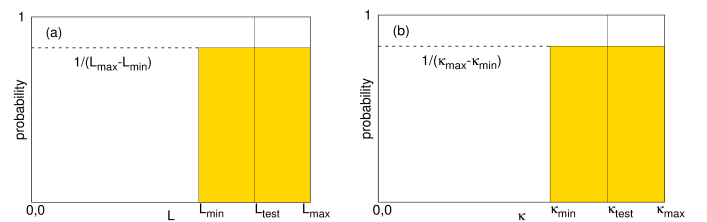}
\caption{The distributions of the system sizes and disorder strengths used in the (imperfect) training of the 
supervised learning algorithm. (a) The distributions of the system sizes are in the training set are randomly
chosen between $L_{min}$ and $L_{max}$ with uniform probability, where $L_{max}$ is fixed at 10000, the threshold distribution
is Weibull, with a shape parameter $\kappa=3.0$. For the testing set, the system size is fixed at $L_{test}=(L_{max}+L_{min})/2$ and $\kappa=3.0$.
(b) The distribution of the values of the shape parameter $\kappa$ in the Weibull threshold distribution of the failure thresholds
of the fibers in the training set. The system size is fixed at 10000 and $\kappa_{test}=(\kappa_{max}+\kappa_{min})/2$ for the testing set.}
\label{fig_2}
\end{figure}
\begin{figure*}
\includegraphics[width=16cm]{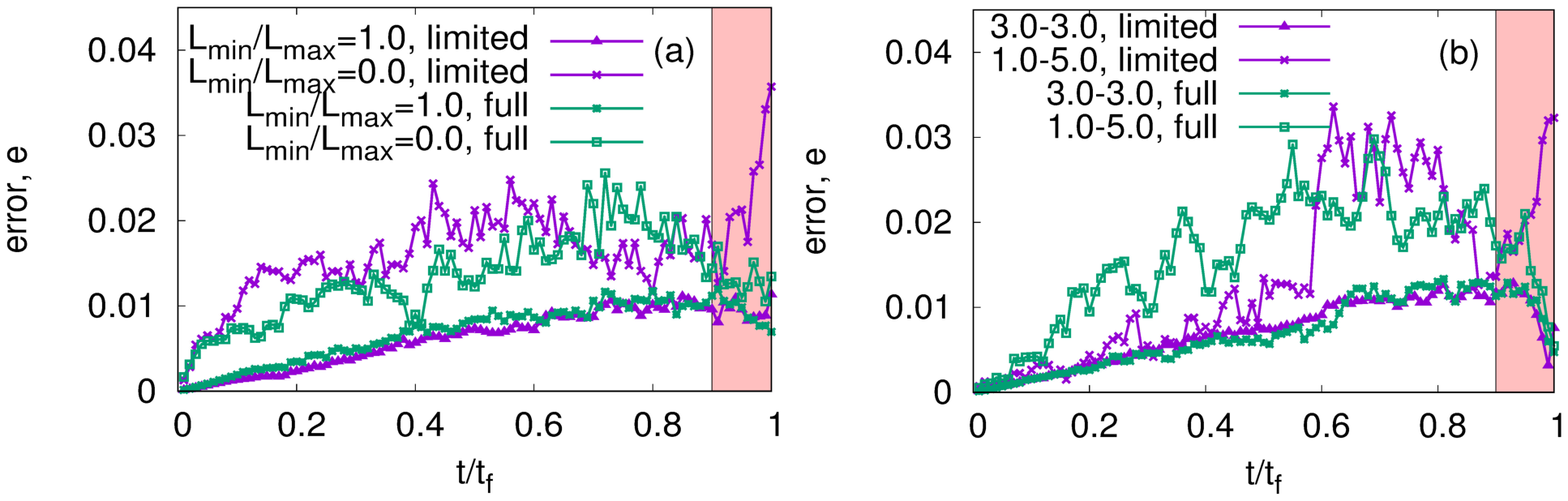}
\caption{The time variations of error $e(t)=\frac{|t_p^{(t)}-t_a^{(t)}|}{t_a^{(t)}}$ where $t$ varies from 0 to $t_f$ are plotted for Case I and Case II (see Fig. \ref{first_fig}). (a) The error series are plotted for $L_{min}/L_{max}=0$ and $L_{min}/L_{max}=1$ in the cases when the training sets did not contain the inequality measures $g,h,k$ (limited training) and when they did contain the inequality measures (full training). (b) The error series are plotted for $(\kappa_{max},\kappa_{min})=(1,5)$ and 
$(\kappa_{max},\kappa_{min})=(3,3)$ in the cases when the training sets did not contain the inequality measures (limited training) and when they did contain the inequality measures (full training). In both (a) and (b), the error values in the last 10\% of time (shared region) jumps up to a higher value when the training sets contain a distribution of system sizes or disorder strengths and did not include the inequality measures. This is a quantitative demonstration of the importance of the inequality measures in predicting failure times when the training sets are imperfect. When the training set samples are all of same size and with same disorder strengths i.e., $L_{min}/L_{max}=1$ and $\kappa_{max}=\kappa_{min}$, then the errors remain almost unaffected by the exclusion of inequality measures in the training features.}
\label{fig_2.5}
\end{figure*}

\begin{figure}
\includegraphics[width=9cm]{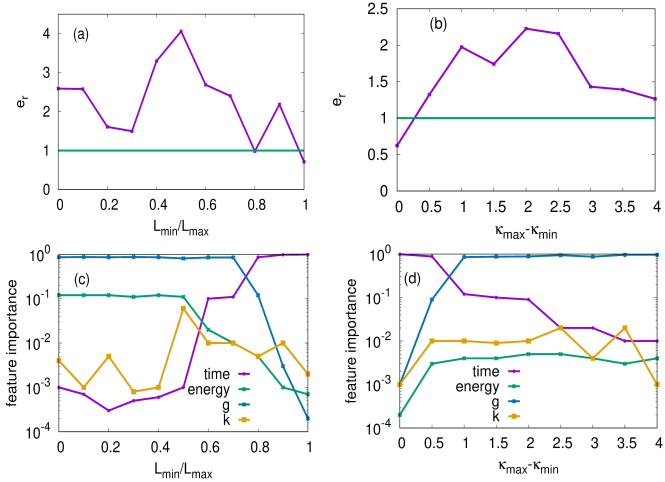}
\caption{The relative errors in predictions and the feature importance for the supervised learning algorithm when the training 
sets are imperfect either in terms of the system sizes or in terms of the disorder strength distributions. (a) and (b) shows the 
ratio of the errors ($e_r=e_1/e_2$) in the predictions when the inequality measures ($h,g,k$) were not used in the training with the case when those
were used in the training. Here $e_x=\sum\limits_{i=0.9t_f}^{t_f}\frac{|t_p^{(i)}-t_a^{(i)}|}{t_f}$, with $x=1,2$ in Case I and Case II respectively
 (see Fig. \ref{fig_2.5}). (c) and (d) show the feature importance values of four major features as functions of imperfections
in the training sets ($L_{min}$ varied from 0 to $L_{max}=10000$ in (c) and $\kappa_{max},\kappa_{min}$ varied from (1,5) to
(3,3) in (d)). For higher imperfections i.e., low values of $L_{min}$ and high values of $\kappa_{max}-\kappa_{min}$, $g$ is the
most important feature, and for lower imperfections, time is the most important feature. Consequently, the relative error in
(a) is greater than 1 for low $L_{min}$ and drops to near 1 for high $L_{min}$, when inequality measures are less important anyway. 
A similar trend is also seen in (b).}
\label{fig_3}
\end{figure}

\begin{figure*}
\includegraphics[width=15cm]{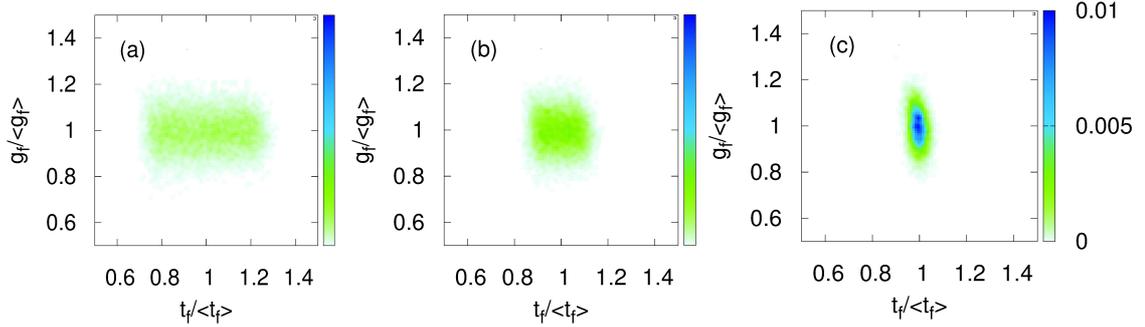}
\caption{The probability distributions $Q(g_f/\langle g_f\rangle,t_f/\langle t_f\rangle)$ are shown (in density map)
for the end-point (just prior to failure) values of $g$ and time. (a-c) show the 
end-point distributions of $g$ an time for $L_{min}/L_{max}=0.6,0.8,1.0$ respectively. As can be seen, the
end-point distribution was narrower along $g$ than $t$ for $L_{min}/L_{max}=0.6$ (a), such that $g$ was a better predictor
in that case (see Fig. \ref{fig_3}(c)). Then in the extreme case $L_{min}/L_{max}=1.0$ (c), the distribution is
narrower along $t$, and it is also the most important feature there.}
\label{fig_4}
\end{figure*}

\begin{figure*}
\includegraphics[width=15cm]{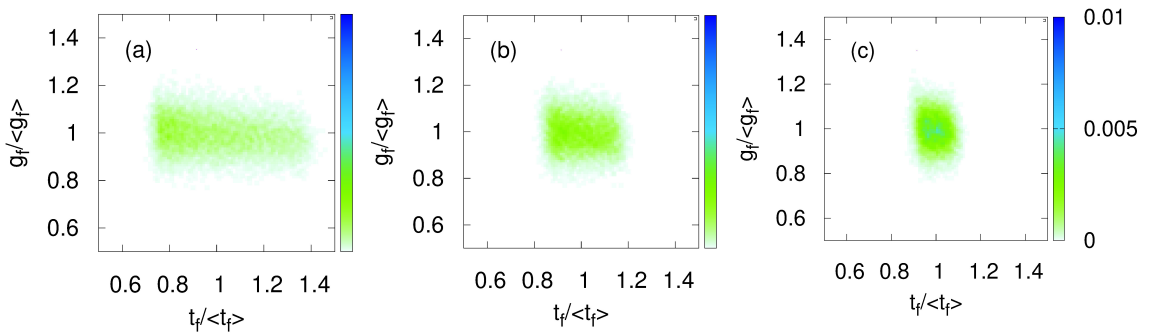}
\caption{The probability distributions $Q(g_f/\langle g_f\rangle,t_f/\langle t_f\rangle)$ are shown (in density map)
for the end-point (just prior to failure) values of $g$ and time. (a-c) show the 
end-point distributions of $g$ an time for $\tau_{max}-\tau_{min}=2.0,1.0,0.5$ respectively.  As can be seen, the
end-point distribution was narrower along $g$ than $t$ for $\tau_{max}-\tau_{min}=2.0$ (a), such that $g$ was a 
better predictor in that case (see Fig. \ref{fig_3}(d)). Then in the case $\tau_{max}-\tau_{min}=0.5$ (c), 
the distribution is
narrower along $t$, and it is also the most important feature there.}
\label{fig_5}
\end{figure*}

\section{Model and Methods}
Fiber bundle model (FBM) was introduce by F. T. Peirce  in 1926 \cite{pierce} in textile engineering for the study of the strength of cotton yarns. 
In FBM, a large number of parallel fibers are attached between two horizontal plates and a load is connected with the lower plate. Each fiber is linear elastic with the same elastic modulus, until its breaking threshold is reached, when it fails irreversibly. The breaking thresholds are different from each other and are drawn randomly from a distribution. The width of this distribution is a measure of the strength of disorder in the model, which is one of the factors governing the mode of failure in the model \cite{failure_mode}. When the attached load is increased, some fibers can break and the load carried by those fibers are then redistributed to the remaining surviving fibers. How the load is redistributed among the surviving fibers is the second factor that governs the failure mode. In general, we redistribute the load of a broken fiber among its $R$ nearest neighbors, when the fibers are arranged in a linear lattice. In this way, $R=1$ gives one extreme limit when the load sharing is fully local i.e., the damage nucleation is preferred. The other extreme is when $R\sim L$, where $L$ is the system size, which the mean-field limit when the load is equally shared between the surviving fibers and the successive damages in the system are governed by the next weakest patch in the system. With no spatial correlation in the disorder of the sample, this limit ($R\sim L$) prefers damages in random locations. On the other hand, a strong disorder in the system can overcome the effect of stress concentration due to localized load sharing i.e., even for a high load the fiber might not break if the failure strength is even higher. So, the interplay between the disorder strength and the load redistribution range lead to different modes of failure in the model with one effect winning over the other for a combination of the values of disorder strength and load redistribution range. 

The most widely studied mode of failure in the model is quasi-brittle, where the disorder strength is moderate and the range of redistribution is global ($R\sim L$). Here, the widely observed scale free distribution of avalanche sizes are reproduced \cite{rmp}.  

It is important to note here that in other threshold activated models of fracture, for example in the random fuse model, the range of load redistribution is irrelevant in the renormalization group sense, and the only mode of failure is nucleation driven in the thermodynamics limit \cite{shk} except in infinite disorder strength \cite{han_za}. In small sizes, however, a critical-like behavior can exist i.e., that shows a scale free distribution of avalanche sizes. Therefore, system size effect is a very important feature in determining failure modes of fracture \cite{shk}.  

The time series of avalanche and the corresponding energy bursts (which are statistically distinct quantities when observed in non mean-field limit \cite{narendra}), hold the key for predicting an imminent catastrophic event in the systems. This is generally true for driven disordered materials \cite{ekhard} and also specifically for the FBM \cite{rmp}. Several rule-based prediction methods have been proposed for the FBM and other models, including the change (lowering) of the exponent of the avalanche size distribution \cite{hemmer}, the linear relation between breaking rate minimum and the failure point \cite{hemmer2}, monitoring elastic energy \cite{hansen1,hansen2} etc. Similar observations have been made in a variety of avalanching disordered systems, both numerically (see e.g., \cite{sim1,sim2,sim3}) and experimentally (see e.g., \cite{exp1,exp2,exp3}). However, experimentally the prediction accuracy are rather limited based on these rule-based approaches \cite{exp3}. We, therefore, need a systematic analysis of the relevant time series features using a supervised machine learning that can extract the information necessary to determine imminent catastrophic events from these multidimensional data. Several such efforts have also been made recently for stick-slip dynamics \cite{leduc} and creep ruptures \cite{sci_rep2} using avalanche time series and its various moments as training features.  

We describe below the method and parameters for simulating the FBM and the details of the features of the time series for avalanche and energies that were used in the supervised learning algorithm. The features used goes beyond the usual time series and its moments and use the inequality measures in such series, which turns out to be very useful in making predictions of imminent breakdown. We then describe the algorithm itself, before moving on to the predictions it makes and analyze the reasons for the importance of different features under different conditions. 

\subsection{Simulations}     
In the simulation, $L$ fibers are taken in a linear array. Each fiber is linear elastic (stress $\propto$ strain) with the same elastic modulus. When the externally applied stress reaches the failure threshold ($\sigma_{th}^i$) of a given ($i$-th) fiber, when it irreversibly fails. The failure thresholds of the individual fibers are randomly selected from a distribution $p(\sigma_{th})$. When a fiber fails, its stress (load) is redistributed equally between $R$ nearest surviving neighbors on both sides ($2R$ fibers in total). 

The simulation is started by applying the minimum load uniformly on all fibers that is necessary to initiate the breaking process i.e., the load needed to break the weakest fiber. When the weakest fiber breaks, its load is redistributed, which may in-turn initiate further breaking and subsequent redistribution. When the system comes to a stable state, i.e. no more fiber is breaking, then the load is increased again uniformly on all fibers until the next fiber breaks and the dynamics is restarted. An avalanche $S$ is defined as the number of fibers breaking between two successive stable states of the system. The corresponding energy burst is $\sum\limits_{i \in S}\left(\sigma_{th}^{(i)}\right)^2/2$, since the fibers are linear elastic. Physically, the distinction between an avalanche size and the corresponding energy burst is that while the former is the crack area opened due to a breaking event, the latter is the elastic energy released due to such opening. These two are, generally speaking, statistically distinct quantities, as was shown theoretically \cite{roy21,narendra} and experimentally \cite{main_expt}. This process of load increase is continued until a catastrophic avalanche sets in, such that the entire system collapses. 

At the end of this breaking process, we end up with two time series viz., the avalanche sizes and the corresponding energy bursts, starting from the initial to the final loading. Given that the time scales of externally applied loading rate and that of the internal stress redistribution are widely separated, the time index denotes the steps of load increase i.e., redistribution are considered to be instantaneous.  

There are two main parameters in the model. First is the range of load redistribution $R$. The second is the failure threshold distribution function $p(\sigma_{th})$, or more specifically characterization of its width. Here we have used two kinds of threshold distributions (specified in respective results): the Weibull distribution 
\begin{equation}
p(\sigma_{th})=\kappa\sigma_{th}^{\kappa-1}e^{-\sigma_{th}^{\kappa}},
\label{weibull}
\end{equation}
 with $\kappa$ being the shape parameter, which is widely used in extreme statistics; and the power-law distribution with cut-offs 
\begin{equation}
p(\sigma_{th}) \sim 1/\sigma_{th}
\label{powerlaw}
\end{equation}
 in the range $(10^{\beta},10^{-\beta})$. The specific values of $\kappa$ and $\beta$ can explore a wide range of disorder strengths in the model.

\subsection{Machine learning method- Random Forest Algorithm}
We have used a supervised machine learning method- the Random Forest (RF) algorithm for predicting the failure time in the fiber bundle model.
The model is first trained using a set of samples with their corresponding time series of avalanche and energy and various attributes derived from those time series (details below). Then the trained model is used for `testing' i.e., making predictions for a different set of data, which is not seen by the model during training. 

RF is an ensemble algorithm that makes an average prediction from a set of decision-trees. In short, the training data are split in each node of a given tree with a threshold for a particular attribute. Each of these groups is further split using a threshold on a different attributes and so on, until in each leaf (end of the decision tree) the value of the target variable (remaining failure time) is the same or further splitting either does not improve predictions (fixed by variance reduction or maximum depth of trees). Now, the training data for each tree is selected from the full training set using bootstrapping (i.e., say $N$ rows uniformly from $N$ rows but with replacements). Bootstrapping acts as a mitigation strategy against outliers in the training set, as some rows are bound to be repeated. In this way, each tree sees a different randomly selected part of the full training set, giving the ensemble of such trees its name - Random Forest. For a detailed discussion, see ref. \cite{ml_book}.

\subsubsection{Features and parameters used}
As indicated above, the two time series of avalanches and energy bursts are used as features, since these are experimentally accessible. Other than that, we also use some inequality measures for the avalanche time series. Specifically, as mentioned before, the response statistics of the system under quasi-brittle conditions are scale-free i.e., events of all sizes (of course limited by the system size) can happen even though the drive (loading rate) is slow and constant. This is generally true for stick-slip type dynamics (see e.g., \cite{leduc}), where spatio-temporal correlations in the response is a mirror to the interplay of damage nucleation and disorder in the systems. As the resulting time series are often scale-free, few of the avalanches represent the vast majority of the damages in the system, while most avalanches are small and account for a tiny fraction of the damage in the system. As larger avalanches tend to occur towards the vicinity of the catastrophic failures, the inequalities in the avalanches grow with time. Quantitatively, these growing inequalities can be represented by certain indices viz., the Gini coefficient ($g$), the Hirsch index ($h$) and the Kolkata index ($k$), traditionally used in socio-economic inequalities (see e.g., \cite{kolkata}). Very recently these measures were also used in fracture and earthquake statistics (see e.g., 
\cite{soc_fbm,benzion}).

Mathematically, Gini coefficient is twice the area between the Lorenz curve ($L(p)$ is the fraction of the avalanche mass held by the $p$ fraction of avalanches when the avalanches are arranged in ascending order) and the equality line for the avalanche time series (see Appendix), Hirsch index ($h$) value gives the number of avalanches that are at least of the size of that particular value and the Kolkata index ($k$) value says that $1-k$ fraction of events hold $k$ fraction of avalanche-mass, which is a generalization of Pareto's law \cite{pareto} (see Appendix). These inequality measures, unlike the size distribution exponents, do not focus on the extreme limits only, but give an `average' estimate of the abundant size of events. These are, therefore, remarkably stable estimates of imminent catastrophic events \cite{k_pre}. 

In summary, the features used in RF training are (unless otherwise mentioned): (i) time, (ii) avalanche size, (iii) energy burst size, (iv) $g$,(v) $h$, (vi) $k$ and the target variable is the remaining time to failure.       

The parameters used for RF are: the number of decision trees used is 1000, the maximum number of splitting (depth of tree) is 10, the minimum data points needed to split a node is 2 and the minimum number of samples in a leaf node is 1. The remaining parameters are default set of scikit-learn 0.19.1 version. The results are stable under small variations of these parameters.      

\subsubsection{Errors and efficiency estimates}
The outputs of the ML algorithm is the predicted remaining time $t_p$ for each sample at each time step. Its difference from the actual time remaining $t_a$ at that time is a measure of the error in the prediction. This error is compared with the error obtained when some features of the training are not used. Particularly, when the inequality indices ($h$, $g$ and $k$) are not used, the errors can change, underlining the necessity for using those features. For this purpose, the time variations of the fractional errors, defined as: $e(t)=\frac{|t_p^{(t)}-t_a^{(t)}|}{t_a^{(t)}}$ and the error ratio $e_r=e_1/e_2$, where $e_x=\sum\limits_{t=0.9tf}^{t_f}\frac{|t_p^{(t)}-t_a^{(t)}|}{t_a^{(t)}}$ are measures. Here $x=1,2$ depending on whether the inequality measures ($g$, $k$, $h$) were not used (1) or used (2) in the ML training. The sum is taken for the last $10\%$ of the time, since in that case only the two predictions differ significantly. The error ratio is a quantitative estimate of the effectiveness of the inequality measures in the training.  

Finally, to estimate the effectiveness of using the machine learning approach itself was estimated as well. In that case, the ratio $e_{ML}/e_{woML}$ is calculated, where
\begin{equation}
e_{ML/woML}= |t_p^{(t)}-t_a^{(t)}|/t_f,
\label{error_est}
\end{equation}
and $t_p$ for $e_{woML}$ is simply obtained from the average remaining time of the training set. The ratio always start with 1 at $t=0$, since in the beginning of the process, average of the training set is the best possible estimate. The ML algorithm `learns' to make better predictions only through the signals emitted during the breaking process. These are discussed later in the following section.

\begin{figure*}
\includegraphics[width=18cm]{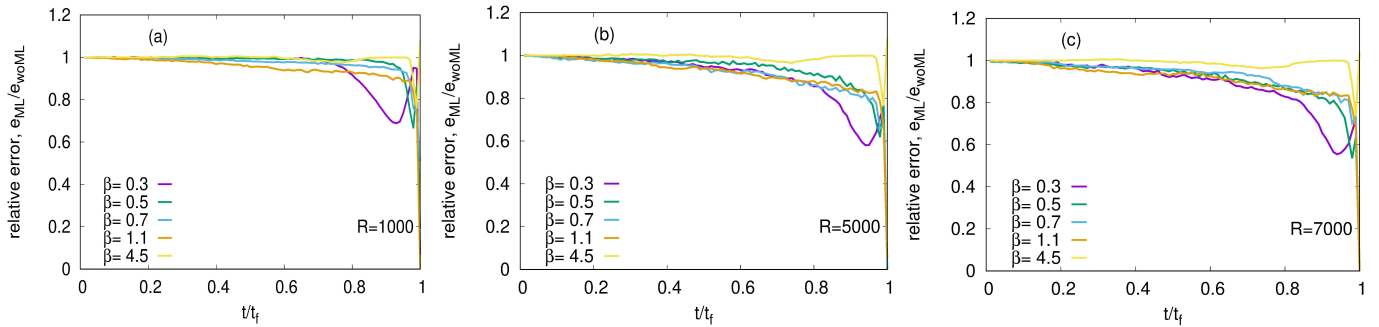}
\caption{The relative errors $e_{ML}/e_{woML}$ are plotted for different interactions ranges $R$ and different strengths of disorder $\beta$. A horizontal line at 1 suggests that the machine learning predictions are simply average prediction of the training set. Whenever the relative errors are below 1, ML predictions are better.  The relative error in predicted failure time for $R=1000, 5000, 7000$ (for left to right) and for various values of disorder $\beta$= 0.3,0.5,0.7,1.1 and 4.5 respectively (for 30000 fibers) are shown. The model is least predictable in the brittle mode (not shown) and best predictable in the quasi-brittle mode ($\beta\approx 0.3$), where avalanche statistics show power-law behavior. Predictability again decreases for high values of $\beta$, where individual fibers fail without showing temporal correlations (see text).}
\label{LLS_preds} 
\end{figure*}

\section{Results}
The fiber bundle model, when used as a model for failure of disordered solids, is usually studied in the mean-field or equal load sharing limit ($R\sim L$ in our notation). The advantages of this limits are: (i) it is analytically tractable, (ii) it reproduces the scale-free statistics of avalanche and energy bursts, usually seen in experiments. The other extreme is the local load sharing limit ($R=1$), where the avalanche size distribution is exponential (nevertheless, the energy burst distribution remains a power-law \cite{narendra}) and the failure is nucleation driven, unless a strong disorder is present \cite{han_za}. 

Each of these situations is crucial for various situations where the FBM is applied and the breakdowns that happen in different ways. Here, we first study the mean-field case, which is the most well-known (quasi-brittle) limit of the model. We focus on the machine learning predictability of the model in this limit and the importance various features, including the physical reasons for such importance. We then report the variation of the effectiveness of the machine learning approach when the mode of the failure in the model changes from percolative (random damage) to nucleation driven failure, due to the interplay of finite $R$ and high disorder strength.

\subsection{Predictability in the Mean field model}
This is the limit when the load of a failed fiber is equally distributed between all surviving neighbors and the strength of disorder in the individual failure thresholds is `moderate' (a narrow distribution would give brittle failure and a very wide distribution would trigger individual fiber breaking at a time or a plastic-like failure). This regions spans from $R=L$ to $R=R_c(L)$ (where $R_c(L)$ scales as some sublinear power of $L$ \cite{failure_mode}) and the disorder strength is moderate. Nevertheless, the simulations presented here follows equal load sharing and the threshold distribution is taken to be Weibull (Eq. (\ref{weibull})). 

\subsubsection{Imperfect training sets and its effect on predictability}
We note that the training set in experimental situations may not be exactly for the same system size or may not have the exact same disorder (impurity) distribution compared to the testing set. In other words, the training set could be imperfect, which can substantially influence the resulting predictions. In particular, we consider a variation in the system size of the training set and a variation of the disorder strength ($\kappa$ values in Eq. (\ref{weibull})) separately. First, a variation in the system size is considered for the training sample, when the disorder strength is held fixed ($\kappa=3.0$). The system sizes are chosen randomly and uniformly for each training sample between the limits $L_{min}$ and $L_{max}$, where $L_{max}=10000$. For the testing samples, $L_{test}=(L_{min}+L_{max})/2$ (see Fig. \ref{fig_2}(a)). Except in the limit where $L_{min}/L_{max}=1$, therefore, the training is imperfect relative to the testing set in terms of system sizes. Then in the second case, the shape parameter (Weibull modulus) $\kappa$ is varied uniformly between the limits $\kappa_{min}$ and $\kappa_{max}$ in the training set (see Fig. \ref{fig_2}(b)), while the system size is kept fixed at $L=10000$. For the corresponding testing set, the $\kappa$ value is fixed at $\kappa_{test}=(\kappa_{max}+\kappa_{min})/2$. 

 These variations in the system sizes and disorder strengths of the training set naturally affect the predictions. It is interesting to note that the errors remain more or less similar in the early stages of predictions irrespective of whether the inequality measures are included in the training features or not (see Fig. \ref{fig_2.5}(a,b)), until towards the end i.e., in the vicinity of the catastrophic failure, where the error shoots up if the inequality measures are not included. This is a quantitative measure of the relevance of the inequality indices when the training sets are imperfect. On the other hand, in the limits when the training set parameters (system size and disorder distribution) match those for the testing set ($L_{min}/L_{max}=1$ and $\kappa_{max}=\kappa_{min}$ respectively), the errors computed for the case with and without inequality indices in feature do not significantly differ (see Fig. \ref{fig_2.5}). This suggests that the inequality indices are particularly important when the training sets are imperfect. We discuss the quantitative measures and the physical reasons for this effect below.

\subsubsection{Feature importance and its reasons}
As indicated above, what physical quantities (features) are vital in prediction are determined by the nature of the training sets. One straightforward way to quantify the relative importance of different features is to measure the feature-importance in the Random Forest algorithm. For all the features of the training data, it gives the relative percentage of importance i.e., which quantities were useful in making the predictions. In particular, how a specific feature decreases the impurity of the sample in each node is measured, then the average over all the decision trees in the Random Forest gives the importance of that specific feature \cite{ml_book}. It is then normalized for all features, giving the relative importance.  

 Fig. \ref{fig_3} shows the error ratios with and without the inequality indices and the importance of the features (four most important ones). In particular, Fig. \ref{fig_3}(c) shows the relative importance of time, energy burst, $g$ and $k$ in predictions, as the ratio $L_{min}/L_{max}$ is varied. As can be seen, for low values of $L_{min}/L_{max}$ (when imperfection in the training set is more), the most important feature by far is $g$. However, as the training set starts resembling the testing set more i.e., for $L_{min}/L_{max}$ closer to 1, time takes over as the most important feature, as the importance of $g$ drops off. Consequently, the error ratio (Fig. \ref{fig_3}(a)) starts from a value higher than 1 and eventually comes close to 1 (since inequality indices are not important anyway for higher values of $L_{min}/L_{max}$, the errors in predictions are not affected by their omission). 

A similar trend is noted for variations in the disorder strength (Figs. \ref{fig_3}(b),(d)). In particular, for relatively pure training sets (low values of $\kappa_{max}-\kappa_{min}$), time is the most important feature, while for higher values of $\kappa_{max}-\kappa_{min}$, $g$ becomes most important by far. Correspondingly, the error ratio (Fig. \ref{fig_3}(b)) starts from a value close to 1 and increases.

It is crucial to explore the reason of such crossover of the most important feature in predictions. It has been recently noted that the inequality indices, $k$ and $g$ in particular, are remarkably stable (near-universal) quantities near self-organized \cite{soc_fbm,soc_all} as well as tuned \cite{k_pre} critical points for a broad class of threshold activated dynamics viz., sand-pile models, driven interfaces and fiber bundles. In critical dynamics of the fiber bundles, the terminal or end-point values of $k(=k_f)$ are almost independent of system sizes and disorder strengths \cite{k_pre} within the quasi-brittle region of failure. This terminal value, therefore, serves as a good indicator of imminent failure. Additionally, it has been noted in a wide variety of systems that $g$ and $k$ are linearly related when their values are not very close to each other \cite{kolkata,soc_all}. This implies that the terminal value for $g=g_f$ can also be a good indicator for imminent catastrophic failure. Now, since $g$ and $k$ values are correlated, importance is given to one ($g$) and not to the other ($k$). In the Appendix, we give the time variations for $g$ and $k$ and show their correlations. In Fig. \ref{fig_4} the  probability distributions $Q(g_f/\langle g_f\rangle,t_f/\langle t_f\rangle)$ of the (scaled) terminal values are shown for different values of $L_{min}/L_{max}$ (0.6, 0.8 and 1.0 from left to right). It can be seen that for $L_{min}/L_{max}=0.6$, the distribution is wider along $t_f$ axis and narrower along the $g_f$ axis, suggesting that $g_f$ is a better indicator here than $t_f$ (which, in fact, is a proxy to the applied load). In other words, using $g$, the training samples can be more effectively split in each node in reducing impurity in that node (as mentioned above).  On the other hand, when $L_{min}/L_{max}=1$, the distribution turns narrower along the $t_f$ axis than the $g_f$ axis, indicating $t_f$ is a better predictor here.  These observations are in line with what is seen in Fig. \ref{fig_3}(c). 

In Fig. \ref{fig_5}, similar plots are shown for different $\kappa_{max}, \kappa_{min}$ values ((2,4), (2.5,3.5), (2.75,3.25) from left to right). Again, the distribution is narrower along the $g_f$ axis for higher imperfection in the training set i.e., $\kappa_{max}-\kappa_{min}=2.0$ and narrower along the $t_f$ axis for $\kappa_{max}-\kappa_{min}=0.5$. The corresponding crossover of importance for $g$ and time are also in line with what is reported in Fig. \ref{fig_3}(d). Similar analysis can be done for the other features, but are not shown here as no other feature assumes a greater importance value than these two in the whole parameter space.

Therefore, it can be concluded that the inequality indices play a vital role in predictions particularly when the training set contains samples that differ in sizes and disorder strengths while maintaining the same mode of failure (quasi-brittle). This is a relevant scenario for almost all experimental cases, where fine tuning of system sizes and disorder strengths are not possible. However, the reason for $g$ and $k$ to have such a stable terminal value, or in fact the reason for the robust nature of the corresponding Lorenz curves (see Appendix), remains a crucial open question.

\subsection{Predictability in the local load sharing model}
Here we simulate the model for local load sharing case, where $R$ is finite and consequently damage nucleation is preferred. As mentioned before, disorder plays an opposing role to localized load sharing in terms of damage nucleation i.e., a strong disorder prefers distributed rather than nucleated damage. To properly capture the effect of moderate to very strong limits of disorder, we take the failure threshold distribution to be power-law with cut-offs (Eq. (\ref{powerlaw})), where $\beta$ determines the strength of disorder. Depending upon the combination of the $\beta,R$ values for a specific system size, the failure mode goes from brittle to quasi-brittle to percolative failure \cite{failure_mode}. This in turn affects the predictability of the model. 

In this case, we fix the system size at $L=30000$ and vary $\beta$ and $R$. The aim here is to quantify the predictability of the model for various different modes and the effectiveness of the machine learning approach itself, with the quasi-brittle mode discussed so far being a limiting case where the redistribution range is large enough such that $R\sim L$. In view of this, the training set is kept `pure' i.e., having the same system size and disorder distribution as the testing set and consequently the inequality indices are not necessary as features in the RF algorithm.

In Fig. \ref{LLS_preds}, we plot the ratios of the errors or the relative error in predictions using machine learning and that without using machine learning (simply the average failure time in the training set. A horizontal line at 1 would indicate the exact same error in predictions as that of the simple average over the training set. As can be seen from Fig. \ref{LLS_preds}, the predictability can also be divided broadly in to three groups.  In the brittle mode, there is no time series as such, given that the whole system collapse as soon as the weakest fiber is broken. Therefore, the ML algorithm can not be used here. Therefore, the best possible prediction in this range is simply the average time of failure of the training set (the time taken by the load to increase to the limit of the weakest fiber). On the other hand, in the limit of percolative failure, there is no temporal correlation in the failure time series, given that each fiber breaks individually. There is no spatial correlation either in the long time, which is, however, not used anyway in the ML algorithm. This implies that the predictability in the percolative mode again is that of the average of the training set. This is verified by the ML predictions for the case of the highest $\beta$ value (4.5), which gives a horizontal line. 
 
In between these two limiting cases, the improvement of predictability is the best when the failure mode just enters the quasi-static regime ($\beta\approx 0.3$ \cite{failure_mode}). This is the regime where the precursory events take place (as opposed to the brittle mode) and also the disorder strength is not to high so as to introduce much sample to sample fluctuation. 
If the disorder strength is increased further than the boundary of the quasi-static mode but still much lower than the percolative mode, then the predictability is better than the percolative and brittle modes but still worse than the boundary of the quasi-brittle regime.  For different values of the range $R$, there is only a rather weak dependence on the disorder strength.

\section{Discussions and conclusions}
Predicting imminent failure in driven disordered systems is a long standing problem in physics, engineering and earth sciences. When external stress is applied on a disordered material, it progresses through avalanching dynamics towards the catastrophic breakdown point. The time-series of these avalanches, reflecting an interplay between damage nucleation and disorder in the system, hold the key of imminent breakdown in their spatio-temporal characteristics \cite{ekhard}. A brittle rupture initiated by damage nucleation and a plastic-like percolative damage, where stress concentration and disorder strengths are dominating respectively, failure points are less predictable (see e.g., \cite{yuta,failure_mode}), since the temporal and spatial correlations are completely absent in these limiting cases.  When the effects of damage nucleation and disorder strength compete, the spatio-temporal correlations are preserved in the avalanches, that show power-law size distribution. In such cases, finding statistical regularities in the avalanche time-series can help in making useful predictions of failure time. As mentioned before, several rule-based approaches have been developed in making such predictions (lowering of size distribution exponent, breaking rate minimum, peak of rate of change in elastic energy etc.) in numerical, analytical and experimental studies. 

Recently, it has also been observed that quantitative measures of inequality in the time series of avalanches also serve as a useful indicators of catastrophic breakdown.
Specifically, the Gini ($g$) \cite{gini} and Kolkata ($k$) \cite{kolkata} indices indicate a characteristic fixed point of the Lorenz curve ($L(p)$: denoting the fraction of avalanche mass in the smallest $p$ fraction of avalanches), that balances the relative abundance of small size avalanches and their weak contribution to the damage with that of the scarcity of very large avalanches that dominate near the catastrophic failure point. Consequently, these fixed point values are often robust \cite{k_pre} -- nearly independent of system sizes and disorder strengths. 

The analysis of the time series of avalanches and energies, along with the inequality measures mentioned above, can therefore extract the relevant information regarding the proximity to failure for a system in a non-invasive manner. We have used the fiber bundle model as a model system for disordered materials. While each fiber is linear elastic with the same elastic modulus, the failure thresholds are chosen from a probability distribution function. Furthermore, as is usual for realistic experimental scenarios, the system sizes and impurity distribution (failure threshold distribution for the model) can also change from sample to sample. Each of these samples, when driven to their respective failure points, will give a series of values in their response statistics (avalanches, energies and the inequalities in avalanches), that will mirror the `health' of the sample i.e., the proximity to catastrophic failure. The statistical regularities in these series are extracted to predict the imminent failure in a sample.

 To do this, we have used a supervised machine learning approach. 
The advantage of a machine learning approach is that it systematically extract statistical regularities from data. 
In this work, we have used the Random Forest algorithm, which is an ensemble of decision trees.
Along with the loading time, avalanche and energy series, we have used the above mentioned inequality indices as features and the target variable is the remaining failure time.
 In each node of a decision tree, the training samples are split using some threshold value of an attribute (a feature) in such a way that the split groups are similar i.e., have similar remaining failure times in this case. A particular feature is given importance when it can better split the sample i.e., can improve the purity of the subgroups. 

It is observed that when the training samples are imperfect i.e., contains samples of different sizes or disorder strengths, the inequality indices, being largely insensitive to these changes, are important predictors for the algorithm (see Fig. \ref{fig_3}). The contribution of these indices in making the predictions are highlighted when such predictions are compared with the same data set but without using the inequality indices (see Fig. \ref{fig_2.5}). It is clear, therefore, that particularly for large imperfections in the training sets, inequality indices play a crucial role in predictions. 
On the other hand, for a training set that contain nominally similar samples i.e., of same system sizes and disorder strengths, then the loading time (a proxy to applied load) overtakes the inequality indices in terms of relative importance.

Finally, when the disorder strengths ($\beta$ see Eq. (\ref{powerlaw})) and stress redistribution range ($R$) are varied, the failure mode of the model goes from brittle (for narrow disorder) to percolative (very wide disorder) through a quasi-brittle region (moderate disorder). As mentioned above, the `learning' process of the algorithm is only useful when there is sufficient signal in terms of the avalanche series and it is only then that the algorithm can make better predictions compared to a simple average over the training set (see Fig. \ref{LLS_preds}). For very low disorder, the failure mode is brittle, hence no learning is possible and on the other extreme with very high disorder, the failures occur with individual breaking where the learning is inadequate. Consequently, in this case, the machine learning approach is most effective in the quasi-brittle failure mode that retains the temporal correlations in the avalanche statistics. 

In conclusion, a systematic analysis of the avalanche and energy time series through supervised machine learning in the fiber bundle model for materials failure can predict imminent catastrophic failure in the model. In the realistic scenarios where training samples are different in size and impurity strength distributions, inequality indices of the avalanche statistics play a vital role in making those predictions. The model is most predictable in the quasi-brittle mode of failure.

\section*{Appendix: Inequality measures for the avalanche time series: Gini ($g$) and Kolkata ($k$) indices}

The avalanche time series in the quasi-brittle region of the FBM does not have a scale in its sizes i.e., it is a power-law distribution $P(S)\sim S^{-5/2}$. This means that large avalanches are few and small avalanches are plenty. The exponent value, of course, is a universal quantity irrespective of system sizes and disorder distributions (so long as it is moderately strong). But the time needed for failure i.e., the critical load is not universal but strongly depends upon the details of the system. Recently it has been shown \cite{k_pre} that some time dependent quantities that reach seemingly universal values near the breakdown point, can be constructed from the measure of inequality between the avalanche sizes. Historically, some of it are over a century old but used mainly in estimating economic inequality (e.g., Gini index \cite{gini}), and some of it are newer (the Hirsch index \cite{hindex} and the Kolkata index \cite{kolkata}). Monitoring those values, therefore, are useful indicators of the approaching failure points. In conjunction with other attributes, they can play a vital role in the supervised learning approach, as was outlined in the main text. 

In this appendix, we describe how those quantities are defined and their time variation and the near-universal terminal values at the critical (breakdown) points are approached in the fiber bundle model.

\begin{figure}
\includegraphics[width=9cm]{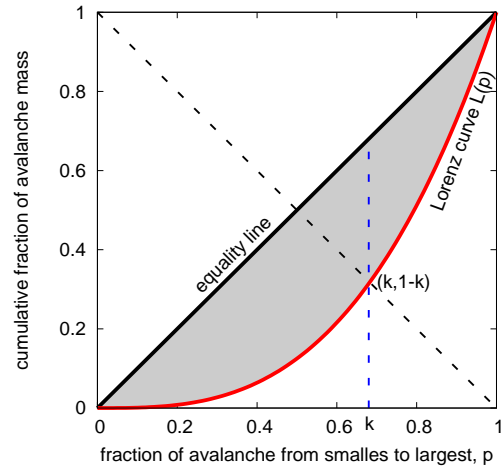}
\caption{A schematic diagram of the Lorenz function $L(p)$ is shown, where $L(p)$ denotes the cumulative fraction of the
avalanche mass contained in the smallest $p$ fraction of avalanches. If all avalanches were equal, this would be a diagonal
straight line, called the equality line. The area between the equality line and the Lorenz curve (shaded area), therefore, is a measure
of the inequality in the avalanche sizes. Two quantitative measures of such inequality are extracted from here, the ratio
of the shaded area and that under the equality line (Gini index, $g$) and the crossing point of the opposite diagonal --
from (0,1) to (1,0), shown in dashed line, and the Lorenz curve, giving the Kolkata index, $k$. $1-k$ fraction of avalanches
contain $k$ fraction of the cumulative avalanche mass.}
\label{fig_s1}
\end{figure}

\begin{figure}
\includegraphics[width=9cm]{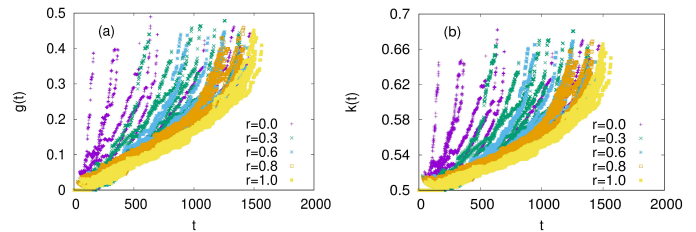}
\caption{The time variations of (a) $g(t)$ and (b) $k(t)$ are shown when individual
samples are of sizes between $L_{min}$ and $L_{max}$ with uniform probability, where $r=L_{min}/L_{max}$ and 
10 time series are shown for each value of $r$. While the failure times are vastly 
different, the terminal values of $g=g_f$ and $k=k_f$ are narrowly distributed. The
failure thresholds are taken from Weibull distribution with a shape parameter value $\kappa=3$ and $L_{max}=10000$.}
\label{fig_s2}
\end{figure}
\begin{figure}
\includegraphics[width=9cm]{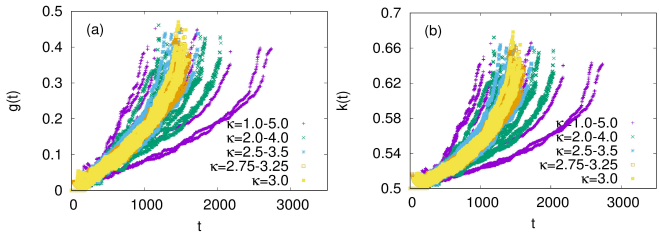}
\caption{The time variations of (a) $g(t)$ and (b) $k(t)$ are shown when individual
samples are of different disorder strengths -- Weibull threshold distributions with 
shape parameters distributed uniformly between $\kappa_{min}$ and $\kappa_{max}$, with 10
samples for each $\kappa_{min},\kappa_{max}$ pair. The system sizes are 1000 always. Again, the 
failure times are vastly different, but the terminal values of $g=g_f$ and $k=k_f$ are narrowly 
distributed.} 
\label{fig_s3}
\end{figure}

\begin{figure}
\includegraphics[width=9cm]{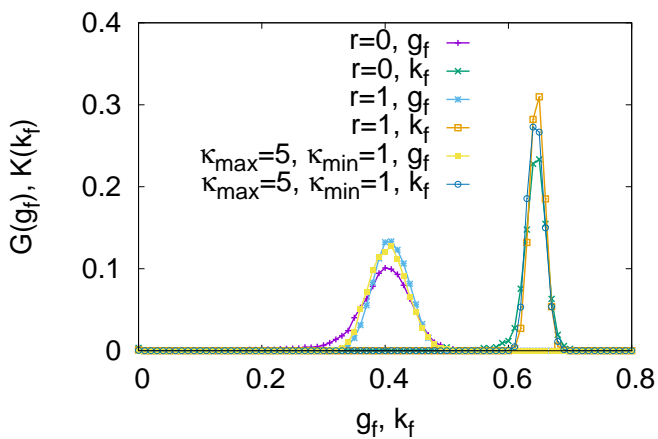}
\caption{The probability distributions of the terminal values of $g=g_f$ and $k=k_f$ are shown for extreme values of $r=L_{min}/L_{max}$ and 
$\kappa_{max}-\kappa_{min}$. The distributions show peak near $\langle g_f\rangle=0.41\pm 0.04$ and $\langle k_f\rangle=0.64\pm 0.02$. The averages are done over 10000 ensembles.} 
\label{fig_s3.5}
\end{figure}

\begin{figure}
\includegraphics[width=9cm]{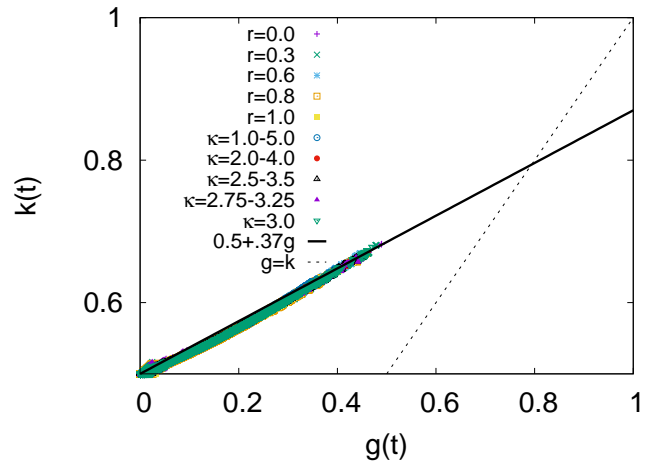}
\caption{The plot of $k(t)$ versus $g(t)$ are shown for various values of $r=L_{min}/L_{max}$ for
fixed $\kappa=3.0$ and various values of $\kappa_{max}-\kappa_{min}$ for fixed system size 10000. 
In all cases, the variations closely follow a linear relation, as was observed in various socio-economic 
systems \cite{kolkata}. The dotted line shows $g=k$ values.}
\label{fig_s4}
\end{figure}

\subsection*{The Lorenz function, Gini ($g$) and Kolkata ($k$) indices}
Upto a given time in the simulation of the FBM, the time series of the avalanches can be arranged in the ascending order. Then the Lorenz function $L(p,t)$ given the cumulative fraction of the avalanche mass (sum of all avalanche sizes) held by $p$ fraction of the smallest avalanches upto time $t$. Note that if all avalanches were of equal sizes, then it would be a straight line starting from origin and increasing upto 1. This is called the equality line (see Fig. \ref{fig_s1}). Since the avalanches are, in general, not of equal sizes, the Lorenz function in non-linear, staying below the equality line and monotonically increasing, with the constraints that $L(0,t)=0$ and $L(1,t)=1$. The area in between the equality line and the Lorenz function, therefore, is a measure of the inequality in the avalanche sizes (the shaded area in Fig. \ref{fig_s1}). The ratio of this area and that under the equality line ($1/2$ by construction) is called the Gini index $g$. 

On the other hand, the ordinate value of the crossing point of the opposite diagonal (straight line between (0,1) to (1,0)), given the value of the Kolkata index $k$, which estimates the fraction $1-k$ of the avalanches that collectively account for the fraction $k$ of the total avalanche mass upto that time. It is a generalization of the Pareto's law \cite{pareto} that says about 80\% of `attempts' account for 20\% of `successes'. It was previously noted that in the case of breakdown in the FBM, at the terminal point $t=t_f$, $k$ approaches a value close to $0.62\pm 0.03$ \cite{k_pre}, irrespective of disorder strengths and system sizes. Figs. \ref{fig_s2}\ref{fig_s3} show the time variations of $g(t)$ and $k(t)$ for different disorder strengths ($\kappa_{max}-\kappa_{min}$ values) and system sizes ($L_{min}/L_{max}$ values). Although the failure
times are strongly dependent on the system sizes and disorder strengths, the terminal values of $g=g_f$ and $k=k_f$ are narrowly distributed (see Fig. \ref{fig_s3.5}). These indices are, therefore, useful indicators of imminent failure (see main text).

It was also noted elsewhere that for a broad class of systems, predominately of socio-economic nature, the early time variations of $g(t)$ and $k(t)$ follow a linear relation $k(t)=1/2+0.37g(t)$ \cite{kolkata,soc_all}. This relation is empirical and seen in data. It turns out that such linearity is also observed in the simulations for the FBM (see Fig. \ref{fig_s4}). As a consequence, (i) the terminal value for $g=g_f$ also approaches a near-universal number ($\approx 0.4$), and (ii) due to the strong correlations between $g(t)$ and $k(t)$, only one of those gets importance in the supervised learning. 

\subsection*{The Hirsch index ($h$)}     
Finally, the Hirsch index $h$, usually measured in the case of citations of individual researchers, takes a value $h$ when there are atleast $h$ avalanches of size $h$ of higher. The terminal value of $h=h_f$, however, is known to depend on the system size as $\propto \sqrt{L}/log(L)$ \cite{k_pre} that makes its predictive power rather limited (as can be seen in the low feature importance values of $h$ in Fig. \ref{fig_3}).

\section*{Acknowledgements}
 We thank Bikas K. Chakrabarti for his comments and suggestions on this manuscript.

\end{document}